\documentclass[aps,prd,onecolumn,groupedaddress,showpacs,nofootinbib,amssymb]{revtex4}
\usepackage[dvips]{graphicx}
\usepackage{amssymb}
\usepackage{amsmath}
\usepackage{graphicx,,color}
\usepackage{amsfonts}
\usepackage{bm}
\usepackage{cancel}
\usepackage{comment}

\newcommand\be{\begin{equation}}
\newcommand\ee{\end{equation}}

\allowdisplaybreaks[4]

\begin{document}

\title{Exponential Inflation with $F(R)$ Gravity}
\author{V.K. Oikonomou,$^{1,2,3}$\,\thanks{v.k.oikonomou1979@gmail.com}}
\affiliation{
$^{1)}$ Department of Physics, Aristotle University of Thessaloniki, Thessaloniki 54124, Greece\\
$^{2)}$ Laboratory for Theoretical Cosmology, Tomsk State University
of Control Systems
and Radioelectronics, 634050 Tomsk, Russia (TUSUR)\\
$^{3)}$ Tomsk State Pedagogical University, 634061 Tomsk, Russia\\
}

\tolerance=5000

\begin{abstract}
In this paper we shall consider an exponential inflationary model in
the context of vacuum $F(R)$ gravity. By using well-known
reconstruction techniques, we shall investigate which $F(R)$ gravity
can realize the exponential inflation scenario at leading order in
terms of the scalar curvature, and we shall calculate the slow-roll
indices and the corresponding observational indices, in the context
of slow-roll inflation. We also provide some general formulas of the
slow-roll and the corresponding observational indices in terms of
the $e$-foldings number. In addition, for the calculation of the
slow-roll and of the observational indices, we shall consider quite
general formulas for which the assumption that all the slow-roll
indices are much smaller than unity, is not necessary to hold true.
Finally, we investigate the phenomenological viability of the model
by comparing it with the latest Planck and BICEP2/Keck-Array
observational data. As we demonstrate, the model is compatible with
the current observational data for a wide range of the free
parameters of the model.
\end{abstract}

\pacs{04.50.Kd, 95.36.+x, 98.80.-k, 98.80.Cq,11.25.-w}

\maketitle

\section{Introduction}

In modern theoretical cosmology there are two widely popular
scenarios that describe in a consistent way the primordial
evolution, the inflationary scenario
\cite{Linde:2007fr,Gorbunov:2011zzc,Lyth:1998xn} and bouncing
cosmology
\cite{reviews1,Brandenberger:2012zb,Brandenberger:2016vhg,Battefeld:2014uga,Novello:2008ra,Cai:2014bea,deHaro:2015wda,Lehners:2011kr,Lehners:2008vx,
Cheung:2016wik,Cai:2016hea}. Both scenarios are quite appealing,
solving most of the shortcomings of the standard Big Bang cosmology,
however there are still many theoretical challenges to address. In
all cases, the theoretical models have to be eventually confronted
with the current observational data coming from the Planck
collaboration \cite{Ade:2015lrj} and the BICEP2/Keck-Array
\cite{Array:2015xqh}. Many models of modified gravity
\cite{reviews1,reviews2,reviews4,reviews5} that describe inflation or
bouncing cosmology, in various theoretical contexts, remain valid,
since the confrontation with the observational data validates their
consistency. For example, in the context of $F(R)$ gravity and modified gravity in general, it is
possible to provide a viable cosmological evolution for various
inflationary scenarios
\cite{Noh:2001ia,Barrow:1988xh,Liddle:1994dx,Hwang:2001qk,Hwang:2001pu,Hwang:1990re,Nojiri:2003ft,Ferraro:2006jd,Nojiri:2007cq,Huang:2013hsb,Hwang:1995bv,Artymowski:2014gea,Brooker:2016oqa,Sebastiani:2015kfa,Odintsov:2016plw,Oikonomou:2015qfh,Odintsov:2015gba}, and also
for bouncing cosmology \cite{Odintsov:2015zza,Odintsov:2015ynk}.
With regard to the inflationary scenario, the standard approach is
to use a slow-rolling scalar field, and an epitome of a large class
of viable scalar-tensor cosmological models is offered by the
$\alpha$-attractor models
\cite{Kallosh:2013hoa,Ferrara:2013rsa,Kallosh:2013yoa}, see also
\cite{Odintsov:2016vzz,Odintsov:2016jwr}, for an $F(R)$ gravity
description of the $\alpha$-attractors. As was demonstrated in Refs.
\cite{Odintsov:2016vzz,Odintsov:2016jwr}, $F(R)$ gravity offers a
fertile ground for the development of viable inflationary theories,
even in the vacuum case. In this line of research, in this paper we
shall consider an exponential model of inflation, in which the
Hubble rate and the corresponding scale factor have the following
form,
\begin{equation}\label{hubbleratemain1}
H(t)=H_0 e^{-\lambda t},\,\,\,a(t)=C_1 e^{-\frac{H_0 e^{\lambda
(-t)}}{\lambda }}\, ,
\end{equation}
where $H_0$ and $C_1$ are real and positive constants. The model
(\ref{hubbleratemain1}) is not so popular in the context of
scalar-tensor inflationary theories, however it is very similar to
the phantom Little Rip inflationary scenario
\cite{Liu:2012iba,Frampton:2011rh}, in which case the Hubble rate is
$H=H_0e^{\lambda t}$. We need to note that similar models of inflation were studied in Ref. \cite{diegonew}. Our first intention was to investigate if the
phantom Little Rip inflationary scenario can be realized by $F(R)$
gravity, and if the resulting inflationary cosmology is viable. It
turns out that only when $\lambda$ is negative, the $F(R)$ gravity
inflationary solution is viable. It is conceivable that the
effective equation of state parameter
$w_{eff}=-1-\frac{2\dot{H}}{3H^2}$ is not phantom for the evolution
(\ref{hubbleratemain1}), and actually it is $w_{eff}=-1+\frac{2
\lambda  e^{\lambda  t}}{3 H_0}-1$. In the following sections we
shall investigate which $F(R)$ gravity can realize the cosmological
evolution (\ref{hubbleratemain1}) at leading order in the large
curvature limit, which corresponds to the inflationary era. By using
the resulting $F(R)$ gravity, we shall perform a detailed analysis
of the inflationary dynamics, by assuming a slow-roll era evolution.
Finally, we shall confront the resulting inflationary model with the
current observational data and we shall analyze the parameter space
in order to see the range of values of the parameters for which the
viability can be achieved. As we demonstrate, the viability of the
model comes for a wide range of parameters.

This paper is organized as follows: In section II we shall briefly
present some essential features of vacuum $F(R)$ gravity, which are
necessary for the following sections. In section III we present the
inflationary dynamics formalism and we express the slow-roll indices
as functions of the $e$-foldings number $N$. We present in detail
the formulas of the slow-roll indices and of the corresponding
observational indices in terms of $N$, and we consider the most
general case for the approximate functional form of the
observational indices. In section III, we employ a well-known
reconstruction technique, in order to find the $F(R)$ which realizes
the exponential inflationary cosmology. In section IV we analyze in
depth the parameter space and we investigate for which values of the
free parameters, the exponential inflationary model in the context
of $F(R)$ gravity, can be viable. Finally, the conclusions follow in
the end of this paper.

Before we proceed to the presentation of our results, we briefly
present the geometric conventions we shall use in this paper. We
shall consider a flat Friedmann-Robertson-Walker (FRW) spacetime,
with the line element being,
\begin{equation}
\label{metricfrw} ds^2 = - dt^2 + a(t)^2 \sum_{i=1,2,3}
\left(dx^i\right)^2\, ,
\end{equation}
and $a(t)$ denotes the Universe's scale factor. Moreover, we shall
assume that the metric connection is a metric compatible affine
connection, which is torsion-less and symmetric, the Levi-Civita
connection.

\section{Basic Features of $F(R)$ Gravity}

In this section we shall briefly present some basic features of
vacuum $F(R)$ gravity, and for more details on this topic, the
reader is referred to Refs. \cite{reviews1,reviews2,reviews4}. The
4-dimensional $F(R)$ gravity gravitational action is equal to,
\begin{equation}\label{action1dse}
\mathcal{S}=\frac{1}{2\kappa^2}\int \mathrm{d}^4x\sqrt{-g}F(R),
\end{equation}
with $\kappa^2$ being $\kappa^2=8\pi G=\frac{1}{M_p^2}$, $g$ being
the determinant of the background metric, and also $M_p$ stands for
the Planck mass. We shall employ the metric formalism, and upon
variation of the action (\ref{action1dse}) with respect to the
metric tensor $g_{\mu \nu}$, the gravitational equations of motion
become,
\begin{equation}\label{eqnmotion}
F_(R)R_{\mu \nu}(g)-\frac{1}{2}F(R)g_{\mu
\nu}-\nabla_{\mu}\nabla_{\nu}F_R(R)+g_{\mu \nu}\square F_R(R)=0\, ,
\end{equation}
which can be rewritten in the following way,
\begin{align}\label{modifiedeinsteineqns}
R_{\mu \nu}-\frac{1}{2}Rg_{\mu
\nu}=\frac{\kappa^2}{F_R(R)}\Big{(}T_{\mu
\nu}+\frac{1}{\kappa^2}\Big{(}\frac{F(R)-RF_R(R)}{2}g_{\mu
\nu}+\nabla_{\mu}\nabla_{\nu}F_R(R)-g_{\mu \nu}\square
F_R(R)\Big{)}\Big{)}\, ,
\end{align}
where $F_R$  stands for $F_R=\frac{\partial F}{\partial R}$. By
using the FRW metric of Eq. (\ref{metricfrw}), the gravitational
equations of motion take the following form,
\begin{align}
\label{JGRG15} 0 =& -\frac{F(R)}{2} + 3\left(H^2 + \dot H\right)
F_R(R) - 18 \left( 4H^2 \dot H + H \ddot H\right) F_{RR}(R)\, ,\\
\label{Cr4b} 0 =& \frac{F(R)}{2} - \left(\dot H +
3H^2\right)F_{R}(R) + 6 \left( 8H^2 \dot H + 4 {\dot H}^2 + 6 H
\ddot H + \dddot H\right) F_{RR}(R) + 36\left( 4H\dot H + \ddot
H\right)^2 F_{RRR}(R) \, ,
\end{align}
where $F_{RR}$ and $F_{RRR}$ stand for $F_{RR}=\frac{\partial^2
F}{\partial R^2}$ and $F_{RRR}=\frac{\partial^3 F}{\partial R^3}$
respectively, and also $H$ denotes the Hubble rate $H=\dot a/a$. In
addition, the ``dot'' indicates differentiation with respect to the
cosmic time, and in addition the Ricci scalar $R$ for the FRW metric
(\ref{metricfrw}) is equal to $R=12H^2 + 6\dot H$.

\section{Inflationary Dynamics of $F(R)$ Gravity: Formalism}

 In this section we shall present the formalism of the $F(R)$ gravity slow-roll inflationary dynamics. Details on the inflationary dynamics for
 $F(R)$ gravity can be found in Refs. \cite{Noh:2001ia,Hwang:2001qk,Hwang:2001pu}, see also
Refs.~\cite{Odintsov:2016plw,Odintsov:2015gba} for some recent
literature on the subject. The slow-roll indices $\epsilon_i$,
$i=1,...,4$ for a general vacuum slow-roll $F(R)$ gravity are,
\begin{equation}
\label{slowrollgenerarlfrphi}
\epsilon_1=-\frac{\dot{H}}{H^2}\,,\quad \epsilon_2=0\, , \quad
\epsilon_3=\frac{\dot{F}_R}{2HF_R}\, ,\quad
\epsilon_4=\frac{\dot{E}}{2HE}\, ,
\end{equation}
with the function $E$ appearing in Eq.~(\ref{slowrollgenerarlfrphi})
being equal to,
\begin{equation}
\label{epsilonfnction} E=\frac{3\dot{F}_R^2}{2\kappa^2}\, .
\end{equation}
Also, a very useful quantity related to the calculation of the
scalar-to-tensor ratio, is $Q_s$, which is defined as follows,
\begin{equation}
\label{qsfunction} Q_s=\frac{E}{F_RH^2(1+\epsilon_3)^2}\, .
\end{equation}
The calculation of the observational indices for the model at hand
may vary, depending on the values that the slow-roll indices take
during the slow-roll era. In the general case, and if the slow-roll
indices satisfy $\dot{\epsilon}_i\simeq 0$ the spectral index of the
primordial curvature perturbation is
\cite{Noh:2001ia,Hwang:2001qk,Hwang:2001pu},
\begin{equation}
\label{spectralindex1}
n_s=4-2\nu_s\, ,
\end{equation}
with the quantity $\nu_s$ being equal to,
\begin{equation}
\label{nus}
\nu_s=\sqrt{\frac{1}{4}+\frac{(1+\epsilon_1-\epsilon_3+\epsilon_4)
(2-\epsilon_3+\epsilon_4)}{(1-\epsilon_1)^2}}\, .
\end{equation}
In the particular case that $\epsilon_i\ll 1$, the spectral index is
approximately equal to,
\begin{equation}
\label{spectralindex2}
n_s\simeq 1-4\epsilon_1+2\epsilon_3-2\epsilon_4\, .
\end{equation}
The scalar to tensor ration $r$ for a vacuum $F(R)$ gravity is
defined as follows \cite{reviews1},
\begin{equation}
\label{scalartotensor1}
r=\frac{8\kappa^2Q_s}{F_R}\, ,
\end{equation}
where we defined $Q_s$ in Eq.~(\ref{qsfunction}). After some
algebra, in the case  at hand, the scalar-to-tensor ratio reads,
\begin{equation}
\label{scalartotensor2}
r=\frac{48\epsilon_3^2}{(1+\epsilon_3)^2}\, .
\end{equation}
In the particular case that $\epsilon_i\ll 1$, the scalar-to-tensor
is greatly simplified, since $\epsilon_1\simeq -\epsilon_3$ and the
above relation is simplified as follows,
\begin{equation}\label{simplifiedscalartotensor}
r=48\epsilon_1^2\, .
\end{equation}
In the rest of this section we shall investigate the behavior of the
slow-roll indices during the slow-roll era, and in principle one can
use the most appropriate definition of the observational indices we
described above. However, regardless of the choice of the
approximation one can use, with regard to the observational indices,
namely Eqs. (\ref{spectralindex1}) and (\ref{spectralindex2}) for
the spectral index, or Eqs. (\ref{scalartotensor1}) and
(\ref{scalartotensor2}) for the scalar-to-tensor ratio, the
viability of the theory is independent of the choice if the
slow-roll indices satisfy the condition $\epsilon_i\ll 1$,
$i=1,...,4$. So in order to be as accurate as possible, we shall
choose the formally more rigid approach, in which the spectral index
is given by Eq. (\ref{spectralindex1}) and the scalar-to-tensor
ratio is given by Eq. (\ref{scalartotensor2}).

Before proceeding, let us further simplify the slow-roll indices
appearing in Eq. (\ref{slowrollgenerarlfrphi}), and after some
algebra we obtain,
\begin{equation}
\label{frgravityconstantroll}
\epsilon_1=-\frac{\dot{H}}{H^2}\,,\quad \epsilon_2=0\, , \quad
\epsilon_3=\frac{\dot{F}_{RR}}{2HF_R}\left(
24H\dot{H}+\ddot{H}\right)\, ,\quad
\epsilon_4=\frac{F_{RRR}}{HF_R}\dot{R}+\frac{\ddot{R}}{H\dot{R}}\, ,
\end{equation}
with $F_{RR}=\frac{\partial^2 F}{\partial R^2}$ and
$F_{RRR}=\frac{\partial^3 F}{\partial R^3}$. For the purposes of our
analysis, we shall express the above quantities in terms of the
$e$-foldings number $N$, so by using the following differentiation
rules,
\begin{equation}\label{asxet2}
\frac{\mathrm{d}}{\mathrm{d}t}=H\frac{\mathrm{d}}{\mathrm{d}N}\, .
\end{equation}
\begin{equation}\label{asxet1}
\frac{\mathrm{d}^2}{\mathrm{d}t^2}=H^2\frac{\mathrm{d}^2}{\mathrm{d}N^2}+H\frac{\mathrm{d}H}{\mathrm{d}N}\frac{\mathrm{d}}{\mathrm{d}N}\,
,
\end{equation}
the slow-roll indices become,
\begin{align}
\label{slowrollgenerarlfrphi} & \epsilon_1=-\frac{H'(N)}{H(N)},
\,\,\, \epsilon_2=0\, ,
\\
\notag & \epsilon_3=\frac{F_{RRR}}{F_R}\Big{(} \frac{6 H(N)
H''(N)+24 H(N)^2 H'(N)+6 H(N) H'(N)^2}{2 H(N)}\Big{)}^2, \\ \notag &
\epsilon_4=\frac{-H''(N)+\frac{2
H'(N)^2}{H(N)}-\frac{H'(N)^2}{H(N)}}{H(N) \epsilon_1}-3 \epsilon_1+
\frac{F_{RRR}}{F_R}\Big{(}6 H(N) H''(N)+6 H'(N)^2+24 H(N) H'(N)
\Big{)}\, .
\end{align}
Thus if the Hubble rate $H(N)$ is known, and also the $F(R)$ gravity
which generates the evolution  $H(N)$, then, the slow-roll indices
and the corresponding observational indices can be found.

In order to proceed, let us express the exponential Hubble rate of
Eq. (\ref{hubbleratemain1}), as a function of the $e$-foldings
number $N$, so by solving the equation $N=\ln a$ with respect to the
cosmic time and by substituting the result in Eq.
(\ref{hubbleratemain1}), the resulting expression for the Hubble
rate is,
\begin{equation}\label{hubbleratemain2}
H(N)=\lambda  \ln \left(C_1 e^{-N}\right)\, ,
\end{equation}
where $C_1$ is the integration constant appearing in the scale
factor (\ref{hubbleratemain1}). Substituting the resulting Hubble
rate of Eq. (\ref{hubbleratemain2}), in the slow-roll indices of Eq.
(\ref{slowrollgenerarlfrphi}), we obtain,
\begin{align}
\label{slowrollindicesexact} & \epsilon_1=\frac{1}{\ln \left(C_1
e^{-N}\right)}, \,\,\, \epsilon_2=0\, ,
\\
\notag & \epsilon_3= 18 \frac{F_{RRR}}{F_R} \lambda ^5 \left(1-4 \ln
\left(C_1 e^{-N}\right)\right)^2 \ln \left(C_1 e^{-N}\right),
\\ \notag & \epsilon_4=-24 \frac{F_{RRR}}{F_R} \lambda ^2 \ln \left(C_1 e^{-N}\right)-\frac{2}{\ln \left(C_1 e^{-N}\right)}
+6 \frac{F_{RRR}}{F_R} \lambda ^2\, .
\end{align}
Accordingly, the spectral index of the primordial curvature
perturbations $n_s$ appearing in Eq. (\ref{spectralindex1}) reads,
\begin{equation}\label{spectralindexresulting}
n_s=4-3 \sqrt{\mathcal{K}(N)}\, ,
\end{equation}
where the function $\mathcal{K}(N)$ stands for,
\begin{align}\label{spectralindexfinalcase}
\mathcal{K}(N)=\frac{\left(\ln \left(C_1 e^{-N}\right) \left(4
\frac{F_{RRR}}{F_R} \lambda ^2 \ln \left(C_1 e^{-N}\right) \left(24
\lambda ^3 \ln \left(C_1 e^{-N}\right) \left(2 \ln \left(C_1
e^{-N}\right)-1\right)+3 \lambda ^3+4\right)-4 \frac{F_{RRR}}{F_R}
\lambda ^2-1\right)+1\right)^2}{\left(\ln \left(C_1
e^{-N}\right)-1\right)^2}
\end{align}
Accordingly, the scalar-to-tensor ratio reads,
\begin{equation}\label{scalartotensorratiofinal}
r=\frac{15552 \left(\frac{F_{RRR}}{F_R}\right)^2 \lambda ^{10}
\left(1-4 \ln \left(C_1 e^{-N}\right)\right)^4 \ln ^2\left(C_1
e^{-N}\right)}{\left(288 \frac{F_{RRR}}{F_R} \lambda ^5 \ln
^3\left(C_1 e^{-N}\right)-144 \frac{F_{RRR}}{F_R} \lambda ^5 \ln
^2\left(C_1 e^{-N}\right)+18 \frac{F_{RRR}}{F_R} \lambda ^5 \ln
\left(C_1 e^{-N}\right)+1\right)^2}
\end{equation}
Thus what remains now to complete the study, is to find the $F(R)$
gravity that generates the evolution (\ref{hubbleratemain2}). Then
by expressing the Ricci scalar as a function of the $e$-foldings
number $N$, we can find the the term $\frac{F_{RRR}}{F_R}$ appearing
above, and the resulting expressions of the slow-roll indices and
therefore also the observational indices can also be found. This is
the subject of the next section.

\section{Reconstruction of the $F(R)$ Gravity Realizing the Exponential Inflationary Era}

Let us now proceed to find the functional form of the $F(R)$ gravity
which realizes the evolution (\ref{hubbleratemain2}). To this end,
we shall employ the reconstruction technique which was developed in
Ref. \cite{Nojiri:2009kx}. The cosmological equation (\ref{JGRG15}),
can be cast in the following form,
\begin{equation}\label{frwf1}
-18\left ( 4H(t)^2\dot{H}(t)+H(t)\ddot{H}(t)\right )F_{RR}(R)+3\left
(H^2(t)+\dot{H}(t) \right )F_R(R)-\frac{F(R)}{2}=0\, ,
\end{equation}
By using the $e$-foldings number $N$, and also the differentiation
rules of Eqs. (\ref{asxet2}) and (\ref{asxet1}), the Eq.
(\ref{frwf1}) is written as follows,
\begin{align}\label{newfrw1}
& -18\left ( 4H^3(N)H'(N)+H^2(N)(H')^2+H^3(N)H''(N) \right
)F_{RR}(R)
\\ \notag & +3\left (H^2(N)+H(N)H'(N) \right
)F_R(R)-\frac{F(R)}{2}=0\, ,
\end{align}
where the primes this time stand for $H'=\mathrm{d}H/\mathrm{d}N$
and $H''=\mathrm{d}^2H/\mathrm{d}N^2$. We introduce the function
$G(N)=H^2(N)$, and by writing the differential equation
(\ref{newfrw1}) in terms of $G(N)$, we obtain,
 \begin{align}\label{newfrw1modfrom}
& -9G(N(R))\left ( 4G'(N(R))+G''(N(R)) \right )F_{RR}(R) +\left
(3G(N)+\frac{3}{2}G'(N(R)) \right )F_R(R)-\frac{F(R)}{2}=0\, ,
\end{align}
with $G'(N)=\mathrm{d}G(N)/\mathrm{d}N$ and
$G''(N)=\mathrm{d}^2G(N)/\mathrm{d}N^2$. We can also express the
Ricci scalar $R$ as a function of $G(N)$ and it reads,
\begin{equation}\label{riccinrelat}
R=3G'(N)+12G(N)\, .
\end{equation}
Hence, the $F(R)$ gravity which realizes the Hubble rate $H(N)$ can
be found by solving the differential equation
(\ref{newfrw1modfrom}). Accordingly, we can find the quantity
$\frac{F_{RRR}}{F_R}$ in terms of $R$ and by expressing $R$ as a
function of $N$ by using Eq. (\ref{riccinrelat}), we can find the
exact form of the slow-roll indices (\ref{slowrollindicesexact}),
and the observational indices can easily be obtained. In the case at
hand, the function $G(N)$ is,
\begin{equation}\label{functiongn}
G(N)=\left(\lambda  \ln \left(C_1 e^{-N}\right)\right)^2\, ,
\end{equation}
and in effect,  the algebraic equation (\ref{riccinrelat}) is equal
to,
\begin{equation}\label{rdeterminingr}
12 \lambda ^2 \ln ^2\left(C_1 e^{-N}\right)-6 \lambda ^2 \ln
\left(C_1 e^{-N}\right)=R\, .
\end{equation}
By solving the above with respect to the $e$-foldings number $N$, we
obtain the following solution,
\begin{equation}\label{nrfunction}
N(R)=\ln \left(C_1 e^{\frac{\sqrt{\lambda ^4+\frac{4 \lambda ^2
R}{3}}}{4 \lambda ^2}-\frac{1}{4}}\right)\, .
\end{equation}
In order to obtain the $F(R)$ gravity in a closed form, we shall
focus on the large curvature limit, which corresponds to the
slow-roll inflationary era, and thus, by using Eqs.
(\ref{functiongn}) and (\ref{nrfunction}),  the differential
equation appearing in Eq. (\ref{newfrw1modfrom}), in the large $R$
limit becomes,
 \begin{align}\label{newfrw1modfromcaseathand1}
&  \left(-\sqrt{3} \lambda ^{3/2} R^2\right)F_{RR}(R) +\frac{R}{4}
F_R(R)-\frac{F(R)}{2}=0\, .
\end{align}
The differential equation (\ref{newfrw1modfromcaseathand1}) can be
solved analytically, and the solution is,
\begin{equation}\label{franalytic1}
F(R)\simeq C_3 R^{\mu }+C_4 R^{\sigma }\, ,
\end{equation}
where $C_3$ and $C_4$ are integration constants, and also the
parameters $\mu$ and $\sigma$ are defined as follows,
\begin{align}\label{parametersmusigma}
& \mu=-\frac{-12 \sqrt{2} \sqrt(4){3} \lambda ^{3/2}-\sqrt{-432
\lambda ^{3/2}+288 \sqrt{3} \lambda ^3+6 \sqrt{3}}-\sqrt{2}
3^{3/4}}{24 \sqrt{2} \sqrt(4){3} \lambda ^{3/2}}\\ \notag &
\sigma=-\frac{-12 \sqrt{2} \sqrt(4){3} \lambda ^{3/2}+\sqrt{-432
\lambda ^{3/2}+288 \sqrt{3} \lambda ^3+6 \sqrt{3}}-\sqrt{2}
3^{3/4}}{24 \sqrt{2} \sqrt(4){3} \lambda ^{3/2}}\, .
\end{align}
The inflationary $F(R)$ gravities of this type were also studied in
Ref. \cite{Sebastiani:2013eqa}. Having the resulting form of the
$F(R)$ gravity at hand, enables us to calculate the spectral index
of the primordial curvature perturbations and the scalar-to-tensor
ratio, and we shall investigate the behavior of the observational
indices in the next section.

\section{Inflationary Phenomenology and Confrontation with the Observational Data}

Let us now turn our focus on the viability of the $F(R)$ gravity
model (\ref{franalytic1}), which realizes the cosmological evolution
(\ref{hubbleratemain1}), in the large curvature limit. So by using
the functional form of the $F(R)$ gravity (\ref{franalytic1}) and
also by substituting the Ricci scalar as a function of the
$e$-foldings number $N$ from Eq. (\ref{rdeterminingr}), the
observational indices (\ref{spectralindexresulting}) and
(\ref{scalartotensorratiofinal}) can be obtained in closed form. The
parameter space is rich, and it consists of $\lambda$, $C_4$, $C_3$
and $C_1$, so the viability with the Planck and BICEP2/Keck-Array
data can be obtained easily for a wide range of parameters. Before
proceeding to the analysis of the parameter space, let us recall the
observational constraints on the spectral index $n_s$ and the
scalar-to-tensor ratio $r$, coming from the Planck data
\cite{Ade:2015lrj}, which are
\begin{equation}
\label{planckdata} n_s=0.9644\pm 0.0049\, , \quad r<0.10\, ,
\end{equation}
while the BICEP2/Keck-Array data \cite{Array:2015xqh} further
constrain the scalar-to-tensor ratio as follows,
\begin{equation}
\label{scalartotensorbicep2}
r<0.07\, ,
\end{equation}
at $95\%$ confidence level. Also, from Eq. (\ref{planckdata}) it is
obvious that the spectral index can be considered as compatible with
the Planck observations, when it takes values in the interval
$n_s=[0.9595,0.9693]$, so we shall take this into account in our
analysis. Let us use some characteristic examples in order to see
the viability of the model, so for $N=60$, $\lambda=2$,
$C_3=29025.7963C_4$ and $C_1=\mathcal{O}(1)$, the observational indices become,
\begin{equation}
\label{nsr} n_s=0.966\, ,\quad r=0.0260371\, ,
\end{equation}
and both are compatible with the Planck and BICEP2/Keck-Array data.
\begin{figure}[h]
\centering
\includegraphics[width=18pc]{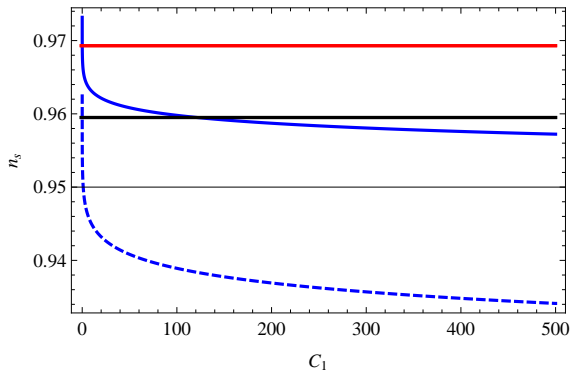}
\caption{The behavior of the spectral index as a function of $C_1$
for $N=60$ (blue thick curve) and for $N=50$ (dashed blue curve),
and with the rest of the parameters being $\lambda=2$,
$C_3=29025.7963C_4$.}\label{plot1}
\end{figure}
There is a large range of the parameters for which the viability of
the model can be achieved, and in order to see this, in
Fig.~\ref{plot1} we plotted the behavior of the spectral index as a
function of  $C_1$ for $N=60$ (blue thick curve) and for $N=50$
(dashed blue curve), and with the rest of the parameters being
$\lambda=2$, $C_3=29025.7963C_4$. In the plots of Fig. \ref{plot1},
the upper red line corresponds to the value $n_s=0.9693$ and the
lower black curve corresponds to $n_s=0.9595$, which is the allowed
range of $n_s$. As it can be seen in Fig. \ref{plot1}, the viability
is achieved for a large range of values of the parameter $C_1$.
\begin{figure}[h]
\centering
\includegraphics[width=18pc]{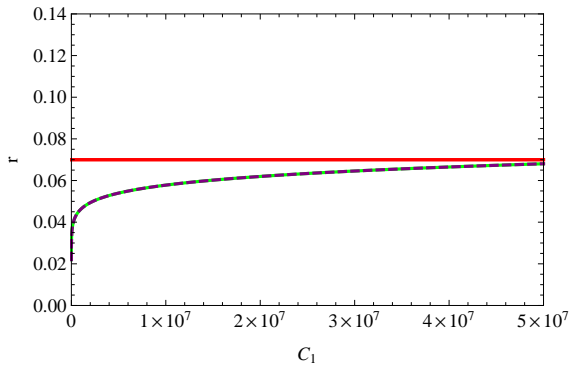}
\caption{The behavior of the scalar-to-tensor ratio, as a function
of the parameter $C_1$, with the green curve corresponding to $N=60$
and the dashed purple corresponding to $N=50$. The upper red line
corresponds to the BICEP2/Keck-Array upper limit $r=0.07$.
}\label{plot2}
\end{figure}
For the same range and values of the parameters, in Fig. \ref{plot2}
we plot the behavior of the scalar-to-tensor ratio, as a function of
the parameter $C_1$, with the green curve corresponding to $N=60$
and the dashed purple corresponding to $N=50$. The upper red line
corresponds the the BICEP2/Keck-Array upper limit $r=0.07$. As it
can be seen Fig. \ref{plot2}, the viability can be achieved for a
large range of the parameter $C_1$, as in the case of the spectral
index. The same applies if other parameters are used, but we omit
for brevity.

Before closing it is worth discussing the limiting values of the parameters $\mu$ and $\sigma$ appearing in Eq. (\ref{parametersmusigma}), as functions of $\lambda$. Particularly, in order for these parameters to be real, the parameter $\lambda$ must be chosen in the following ranges,
\begin{equation}\label{rangeoflambda}
0\leq \lambda \leq \frac{1}{2 \sqrt{3}\,{6 \left(17+12 \sqrt{2}\right)}}\,\,(\simeq 0.08495),\,\,\,\lambda \geq \frac{\sqrt{3}}{2} {\frac{1}{6} \left(17+12 \sqrt{2}\right)}\,\, (\simeq 0.899)\, .
\end{equation}
Then it is easy to see how the resulting $F(R)$ gravity behaves for the limiting values of the parameter $\lambda$. For example if $\lambda$ is chosen to be very small, say $\lambda\simeq \mathcal{O}( 10^{-6})$, then the parameter $\sigma$ is approximately equal to $\sigma\simeq 2$, while $\mu$ takes large values, that is $\mu\simeq 1.44338\times 10^{8}$, so the resulting $F(R)$ gravity is,
\begin{equation}\label{frgravitylimiting123}
F(R)\simeq C_3 R^{1.44338\times 10^{8} }+C_4 R^{2 }\, ,
\end{equation}
so the subdominant term of the resulting $F(R)$ gravity in this case resembles the Starobinsky model \cite{starob}, however this is just the subdominant term. For $\lambda\simeq \frac{1}{2 \sqrt{3}\,{6 \left(17+12 \sqrt{2}\right)}}$, the $F(R)$ gravity becomes approximately,
\begin{equation}\label{frgravitylimiting1231}
F(R)\simeq C_3 R^{3.45445 }+C_4 R^{3.37509 }\, ,
\end{equation}
so the two terms have similar behavior. The viability of the resulting $F(R)$ gravity can also be checked, and of course it is a different model from the Starobinsky model. Indeed, for $\lambda= \frac{1}{2 \sqrt{3}\,{6 \left(17+12 \sqrt{2}\right)}}$, $C_1=\mathcal{O}(1)$ and $N=60$, if the parameters $C_4$ and $C_3$ satisfy $C_4=-1.6162 C_3$, the spectral index is equal to $n_s=0.966$ and the scalar-to-tensor ratio is $r=0.013$. Now when $\lambda$ is equal to $\lambda=\frac{\sqrt{3}}{2} {\frac{1}{6} \left(17+12 \sqrt{2}\right)}$, the $F(R)$ gravity tends to,
\begin{equation}\label{frgravitylimiting1232}
F(R)\simeq C_3 R^{0.640968 }+C_4 R^{0.528364 }\, ,
\end{equation}
and in this case if $C_1=\mathcal{O}(1)$, $N=60$, and $C_4=-3.94335 C_3$, we get $(n_s,r)=(0.966,0.013)$. Finally, for large values of $\lambda$, the parameter $\mu$ tends to $\mu \to 1$ while the parameter $\sigma$  tends to $\sigma \to 0$. In this case the viability of the theory is questionable though, and can be achieved if the parameters $C_4$ and $C_3$ take abnormally large values. We omit this case since it is not so physically appealing. In conclusion, apart from the case that $\lambda$ is extremely large, when the parameter $\lambda$ satisfies the constraints (\ref{rangeoflambda}), the resulting theory can be compatible with the observational data.

\section{Conclusions}

In this paper we studied an exponential inflationary evolution in
the context of slow-roll vacuum $F(R)$ gravity, and we analyzed the
inflationary dynamics in some detail. Particularly, we expressed the
slow-roll indices in terms of the $e$-foldings number $N$, and we
calculated the spectral index of the primordial curvature
perturbations and also the scalar-to-tensor ratio. For the
calculation, the $F(R)$ gravity which realizes the exponential
inflationary scenario was needed, so by using well-know
reconstruction techniques, we found the leading order functional
form of the $F(R)$ gravity, in the large curvature limit, which
characterizes the inflationary era. As we demonstrated, the
resulting inflationary theory is compatible with both the latest
Planck and BICEP2/Keck-Array data, and the compatibility may be
achieved for a large range of parameter values. Also we need to mention that exponential inflation theories of this type, may be the key element also for the construction of unified models of inflation with dark energy in various frames of $F(R)$ gravity \cite{new1serg,new2serg}.

An issue which we did not addressed is the graceful exit issue. In
this theory, this may be achieved due to possible growing curvature
perturbations, but this task is not easy to tackle in the context of
the exponential inflationary theory. However, due to the fact that
the Hubble rate is a quasi-de Sitter evolution, at leading order in
the cosmic time, then the exit comes as an effect of growing
curvature perturbations, as in Refs. \cite{Bamba:2014jia}. Due to
the complexity of this issue, and in order not to fall into
inconsistencies, we hope to formally address this in a future work.

Also, the $F(G)$
\cite{Bamba:2014mya,Bamba:2009uf,Cognola:2006eg,Oikonomou:2015qha,Makarenko:2016jsy}, $F(T)$ gravity \cite{Cai:2015emx,Cai:2011tc} and higher order gravity \cite{Capozziello:2007vd,Clifton:2006kc,Chakraborty:2016ydo} realization of this
theory may also be a subject of future work, which we hope to
address in due time.

\end{document}